 \documentclass[prl,aps,twocolumn,showpacs]{revtex4}
\usepackage[dvips]{graphicx}
\usepackage{dcolumn}
\newcommand{\be}{\begin{equation}}
\newcommand{\ee}{\end{equation}}
\newcommand{\bea}{\begin{eqnarray}}
\newcommand{\eea}{\end{eqnarray}}
\newcommand{\nn}{ \nonumber}

\begin{document}
\topmargin=-20mm

\title{ On the Theory of a New Maximum Observed in dc Transport   in Modulated Quantum Hall Systems Near $ \nu = 1/2$} 

\author{Nataliya A. Zimbovskaya and Joseph L. Birman}

\affiliation{Department of Physics, The City College of CUNY, New York, NY, 10031, USA} 
 

\begin{abstract}
 We propose a theory for the new maximum recently observed by Willett et al [Phys. Rev. Lett. {\bf 83},  2624  (1999)] in the longitudinal magnetoresistance of a weakly modulated 
two-dimensional  electron gas (2DEG)  near filling factor $ \nu =1/2$ for the current driven along the modulation lines. The maximum is superimposed upon a new  resonance structure. This occurs due to the  geometric resonance of composite fermion
cyclotron orbits with the period of modulation of the effective magnetic field. We propose here a semiquantitative theory of the dc magnetotransport in a modulated 2DEG near one half filling. Our analysis is based on the Boltzmann transport equation and it enables us to obtain this maximum which was neither observed nor predicted previously. 
              \end{abstract}

\maketitle

The present work is motivated by the new experimental results of dc transport experiments of Willett et al \cite{1} in a 
two-dimensional electron gas (2DEG) in the fractional quantum 
Hall regime near half filling of the lowest Landau level ($\nu = 1/2$).  The 2DEG was modulated with a small period density modulation applied in one direction. A minimum in the magnetic field dependence of  the dc resistivity component for a current driven {across} the modulation lines, and a maximum in the magnetoresistance corresponding to a current driven {along} the modulation lines, was observed at about $\nu =1/2 $. Both minimum and maximum are superimposed upon a new resonance structure produced by the modulation in the region immediately 
adjacent to $\nu =1/2 $.

Similar structures corresponding to a special kind of geometric resonance (so called Weiss oscillations) were observed in dc transport experiments in both electrostatic and magnetic field modulated 2DEG in low magnetic fields \cite{2,3,4,5}. A theory of these oscillations in modulated 2DEG systems was first developed within the framework of a quantum mechanical approach
\cite{6,7,8}. An equivalent semiclassical approach to the analysis of these phenomena was first proposed by Beenakker \cite{9} who pointed that Weiss oscillations could be explained by means of the guiding center drift of cyclotron orbits of the electrons in the presence of the modulating electric field. The most complete semiclassical consideration of magneto-transport in a modulated 2DEG is presented in Ref. \cite{10} (See also
Ref. \cite{11}).

The minimum in the magnetoresistivity corresponding to a current driven across the modulation lines ($\rho_{\perp}$) at $\nu =1/2 $ was previously observed in the experiments \cite{12}. A theoretical description of this feature of the magnetoresistivity $\rho_{\perp}$ was presented in Refs. \cite{13,14,15}. Here we focus on the maximum in the magnetic field dependence of the magnetoresistivity for a current driven {along} the modulation lines ($\rho_{\parallel}$). Such a maximum was never observed before and it was never predicted in theoretical studies. We develop a semiquantitative theory of the dc magnetotransport which gives a new explanation of this 
maximum observed in the experiments \cite{1}.

 Our work is based on the theory of the quantum Hall system at and near $ \nu = 1/2 $ proposed by Halperin, Lee and Read (HLR)  \cite{16,17}, which corresponds to the physical picture of the electrons decorated by attached quantum flux tubes. These are the relevant quasiparticles of the system -- so called composite fermions (CF). The CFs are charged spinless
fermionic quasiparticles which move in the reduced effective magnetic field $ B_{eff} = B - 4 \pi \hbar c n /e \ (n $ is the electron density) \cite{23}. At $ \nu = 1/2 $ the CFs form a Fermi sea and exhibit a Fermi surface (FS). For the unmodulated 2DEG the CF-FS can be supposed to be a circle in quasimomenta space. Its radius $ p_F $ equals $ \sqrt{4 \pi n \hbar^2}. $

Density modulation influences the CF system through the direct effect of the modulating potential which can deform the CF--FS \cite{22} and through the effect of an additional inhomogeneous magnetic field $ \Delta B \bf (r) $ proportional to the
density modulation $ \Delta n {\bf (r)} \ \big (\Delta B {\bf (r)} = - 4 \pi \hbar c \Delta n {\bf (r)} /e \big) $  \cite{14}. To analyse the effect systematically we have to solve the Boltzmann transport equation for the CF distribution function in the presence of a spatially inhomogeneous disturbance due to the density modulation,  similarly to Refs. \cite{10,15}. When, however, the CFs mean free path $ l $ is larger than the radius of their cyclotron orbit at the effective magnetic field $ R $ and the period of modulation $ \lambda $ we can obtain the desired response functions by means of simplified considerations
based on the works of Beenakker \cite{9} and Gerhardts \cite{11}. 

We start from the Lorentz force equations describing the CF motion along the orbit:
         \begin{equation}
\frac{d p_x}{d t} = - \frac{e}{c} B ({\bf r}) v_y; 
\qquad  \frac{d p_y}{d t} =  \frac{e}{c} B ({\bf r}) v_x, 
                               \end{equation} 
 where $ p_{x,y} $ and $v_{x,y} $ are the components of the CF
quasimomentum and velocity; $ B{\bf (r)} = B_{eff} - 4 \pi \hbar c \Delta  n {\bf (r)} /e.$

We will consider  a single-harmonic sinusoidal density modulation of period $ \lambda = 2 \pi / g $ along the $ ''y'' $ direction: $ \Delta n {\bf (r)} \equiv \Delta n (y) = \Delta n \sin (g y) .$ We assume that the correction term $ \Delta B \bf (r) $ is small compared to $ B_{eff}. $ Under this assumption we can write the CF velocity $ \bf v $ in the form $
{\bf v = v}_0 + \delta \bf v, $ where $ {\bf v}_0 $ is the uniform-field velocity and the correction $ \delta \bf v $ arises due to the inhomogeneity of the magnetic field. For a circular CF-FS we have:  $ v_{x0} = v_F \cos \Omega t ; \ v_{y0} = v_F \sin \Omega t;\  \Delta n (y) \approx \Delta n \sin (gY - g R \cos \Omega t) ,$ where $ v_F $ is the CFs Fermi velocity, $\Omega$  is their cyclotron frequency in the 
effective magnetic field and $ Y $ is the $ ''y''$ coordinate of the guiding center.  Substituting these expressions for $ \bf v $ and $ \Delta n (y) $ into Eq. (1) and keeping only first-order terms we obtain: 
  \[
\frac{d (\delta v_x)}{dt} = - \Omega \delta v_y - \frac{\Delta B}{B_{eff}}
\Omega v_F \sin \Omega t \sin (g Y - g R \cos \Omega t);
         \]
   \be
\frac{d (\delta v_y)}{dt} =  \Omega \delta v_x + \frac{\Delta B}{B_{eff}}
\Omega v_F \cos \Omega t \sin (g Y - g R \cos \Omega t). 
                               \ee

We have to remark here that in the presence of the density modulation the CF's Fermi velocity gets a correction  due to the modulation. To evaluate this correction we calculate the average of $ \Delta n (y)$ over the cyclotron orbit. Expanding the functions $ \cos (gR \cos \Omega t) $ and $ \sin (gR \cos \Omega t) $ in Bessel functions we arrive at the following result for the averaged correction to the inhomogeneous density
modulation: 
         \bea
\big <\Delta n(y)\big > &\equiv& \frac{\Delta n}{2 \pi} \int _0^{2 \pi} \sin
(g Y - gR \cos \psi) d \psi 
   \nn \\   &
= & \Delta n \sin (g Y) J_0 (g R).
                               \eea
 Here $ \psi = \Omega t. $

The result (3) gives a spatially inhomogeneous correction to the chemical potential of the CFs and to their Fermi velocity:
   \[  
\tilde
v_F = v_F \sqrt{1 + \frac{\Delta n}{n} \sin (gY) J_0 (gR )}.
      \]          
  The same holds for the scattering rate of CFs. In periodically modulated  samples the scattering rate can be a function of a coordinates. Within the chosen geometry (the modulation is applied along the $"y"$ direction) 
we can describe a periodic modulation of the relaxation time following  Ref. \cite{10}: 
   \[
 \displaystyle{\frac{1}{\tau(y)}=\frac{1}{\tau}+r(y)} .
      \]

  The correction $r(y)$  
here is proportional to $\displaystyle{{\Delta n(y)}/{n}}$. However we can neglect these corrections to the CF Fermi velocity and their relaxation time in further calculations because the parameter $\displaystyle{({\Delta n}/{n} )
= ({\Delta B}/{B})} $ is much smaller than the parameter
 $\displaystyle{{\Delta B}/{B_{eff}}}$. Thus omitted terms are in order of magnitude smaller than those which are kept.  It is natural to suppose that to the first order in the modulating field the corrections $ \delta v_x $ and $ \delta v_y $ are periodic over the unperturbed cyclotron orbit. This
assumption is   equivalent to that used in Ref. \cite{11}. Under this assumption we can calculate averages of Eq. (2) over the cyclotron orbit. This gives us the following expressions for the components of the velocity of the guiding center drift $ V_x $ and $ V_y $ defined below:
    \bea 
V_x (Y) &=& \big<\delta v_x\big> = - \frac{v_F}{2 \pi} \frac{\Delta B}{B_{eff}} \int_0^{2 \pi} \cos \psi \sin (g Y 
  \nn \\ 
& - & g R \cos \psi) d \psi =
v_F \frac{\Delta B}{B_{eff}} \cos g Y J_1 (g R);
                 \nn \\ 
V_y (Y) & =& \big<\delta v_y\big> = - \frac{v_F}{2 \pi} \frac{\Delta B}{B_{eff}} \int_0^{2 \pi} \sin \psi \sin (g Y
   \nn \\ 
   & - & g R \cos \psi) d \psi 
   = 0.
                               \eea 
  To evaluate  the CF conductivity semiquantitatively we assume that the  $ x $ component of the CF velocity can be written in the form $ v_x (Y) = v_{x 0} + V_x (Y). $ We also assume that the cyclotron frequency $\Omega $ can be replaced by the quantity  $ \Omega (Y) = \Omega + \big<\Delta \Omega (y)\big> $ where $\big<\Delta \Omega (y)\big>$  is the correction to the
cyclotron frequency arising due to the inhomogeneity of the effective magnetic field averaged over the cyclotron orbit:
     \bea
 \Omega (Y) &=& \Omega \left \{1 + \frac{\Delta B }{B_{eff}} \frac{1}{2\pi}
\int_0^{2 \pi} \sin (gY - gR \cos \psi) d \psi \right \} 
      \nn \\    &=& 
     \Omega
\left \{1 + \frac{\Delta B}{B_{eff}} \sin (gY) J_0 (gR) \right \}.
                               \eea 
 We showed before \cite{22} that in the semiquantitative analysis of the magnetotransport in modulated the 2DEG we can use the following approximation for the CF conductivity: 
         \begin{equation}
\sigma_{\alpha \beta}^{cf} \approx \frac{g}{2 \pi} \int_{-\pi
/g}^{\pi /g} \sigma_{\alpha \beta}^{cf} (Y) d Y,
                               \end{equation} 
 where 
         \begin{equation}
\sigma_{\alpha \beta}^{cf} (Y) = \frac{e^2 m_c \tau}{2 \pi \hbar^2} 
\sum_k \frac{v_{k \beta} (Y) v_{- k \beta} (Y)}{1 + i k \Omega
(Y) \tau}.
                               \end{equation} 
 Here $ m_c $ is the CF cyclotron mass; $ \tau $ is the relaxation time; $ v_{k \beta} (Y) $ are the Fourier transforms for the CF velocity components: 
   \bea &&
     v_{kx} = \frac{ v_F}{2} (\delta_{k,1}
+ \delta_{ k,-1}) + V_x (Y) \delta_{k,0}; 
   \nn \\  &&
 v_{ky} = \frac{i v_F}{2} (\delta_{k,1} - \delta_{ k,-1}).\nn
   \eea

 Keeping  terms of the order of $ (\Delta B /B_{eff})^2 $ or larger we obtain the following approximations for the CF conductivity components 
$ \ \displaystyle{\left (\frac{\Delta B}{B_{eff}} \Omega \tau =
\frac{\Delta n}{n} \frac{p_F}{\hbar} l = \frac{\Delta n}{n} k_F l \right ): }$
            \bea     
\sigma_{xx}^{cf} &\approx& \frac{\sigma_0}{1 + (\Omega \tau)^2} \left \{1 + \frac{3}{2} \left (\frac{\Delta n}{n} k_F l \right )^2 \frac{J_0^2 (gR)}{1 + (\Omega \tau)^2} \right \} 
   \nn \\  &+ &
\sigma_0 \left (\frac{\Delta B}{B_{eff}} \right )^2 J_1^2 (g R);
       \\                   
\sigma_{yy}^{cf} &\approx& \frac{\sigma_0}{1 + (\Omega \tau)^2} \left \{1 + \frac{3}{2} \left (\frac{\Delta n}{n} k_F l \right )^2  \frac{J_0^2 (gR)}{1 + (\Omega \tau)^2} \right \}; 
          \nn \\ \\     
 \sigma_{xy}^{cf} &=& - \sigma_{yx}^{cf}
   \nn \\  &
\approx& \frac{\sigma_0 \Omega \tau}{1 + (\Omega \tau)^2} 
\left \{1 + \frac{1}{2} \left (\frac{\Delta n}{n} k_F l \right )^2 \frac{J_0^2 (gR)}{1 + (\Omega \tau)^2} \right \}. \nn \\
                               \eea 
 where $ \sigma_0 = n e^2 l / p_F $ is the CF conductivity in a
homogeneous magnetic field.

The last term in the expression for $ \sigma_{xx}^{cf} $ describes the contribution from CFs diffusing along the $ ''x'' $ direction which arises due to the guiding center drift. To show it we can calculate the corresponding contribution to the diffusion coefficient $ \delta D.$ Following Refs. \cite{9,11} we write: 
         \begin{equation}
\delta D = \tau \frac{g}{2 \pi} \int _{- \pi /g}^{\pi /g} V_x^2 (Y) d Y.
                               \end{equation} 
 This term $ \delta D $ gives the additional contribution to the $ ''x''$ component of the diffusion tensor $ D. $ The latter is connected with the CF conductivity through the Einstein relation $ \sigma_{\alpha \beta}^{cf} = N e^2 D_{\alpha \beta} \  (N $ is the CF density of states).
Substituting Eq. (11) into this relation we obtain the expression for this diffusion correction to $ \sigma_{xx}^{cf} $ which coincides with the last term in Eq. (8).

According to the HLR theory, the 2DEG resistivity tensor $ \rho $ equals: $ \rho = \rho^{cf} + \rho^{cs} $ where $ \rho^{cf} $ is the CF resistivity tensor $ \big (\rho^{cf} = (\sigma^{cf})^{-1} \big ) $ and the contribution $ \rho^{cs} $ arises due to a fictitious magnetic field which originates from the Chern-Simons formulation of the theory. The latter has only off diagonal elements. Hence the diagonal components of the 2 DEG
resistivity tensor coincide with the corresponding components of the CF resistivity tensor $ \rho^{cf}. $ After straightforward calculations we arrive at the result:
         \bea
\rho_{xx} &\approx & \frac{1}{\sigma_0} \left \{1 + \left (\frac{\Delta B}{B_{eff}} \right )^2 \chi_1 (gR) \right \}^{-1};
      \\ 
\rho_{yy} &\approx &\frac{1}{\sigma_0} \left \{1 + \left(\frac{\Delta n}{n} k_F l \right )^2 \frac{\chi_2 (gR)}{1 + (\Delta B/B_{eff})^2 \chi_3 (gR)} \right \}. \nn \\
                               \eea
 Here $ \chi_i (gR) = \alpha_i J_0^2 (gR) + J_1^2 (gR) \; (i = 1,2,3) $ and the coefficients $ \alpha_i $ are given by the expressions:
         \begin{equation}
\alpha_2 = 0;
\quad \alpha_1 = - \frac{1}{2} \,\frac{(\Omega \tau)^2}{1 + (\Omega \tau)^2};
\quad
\alpha_3 =  \frac{(\Omega \tau)^2}{1 + (\Omega \tau)^2} \, .
                               \end{equation} 

When the density modulation is very weak $\displaystyle{\big[({\Delta n}/{n}) k_F l \ll 1 \big]}$ the corrections to the magnetoresistivity are small and we can neglect them. Under this condition the inhomogeneity of the effective magnetic field does not significantly affect dc transport.
For stronger modulation $\displaystyle{\big[({\Delta n}/{n}) k_F l \sim 1 \big]}$ the resistivity component $\rho_{\perp}$ can be significantly changed due to the modulation. Keeping only the greatest correction (in the small parameter $\displaystyle{{\Delta B}/{B_{eff}}}$) we can write for this
component of  the magnetoresistivity the following approximate expression:
      \[ 
\rho_{\perp} \approx\frac{1}{\sigma_{0}}\left \{1+
\left(\frac{\Delta n}{n}k_{F}l\right)^2J_{1}^2(gR)\right\}.
    \]
        
We also see that the resistivity component $ \rho_{\parallel} $ can be  considerably changed due to the modulation: 
          \begin{equation}
\rho_{\parallel}\approx\frac{1}{\sigma_{0}}\left[1+
\left(\frac{\Delta n}{n} 
k_{F}l\right)^2\displaystyle{\frac{1}{(\Omega\tau)^2}}
\left(\frac{1}{2}J_{0}^2(gR)
-J_{1}^2(gR)\right)\right].
           \end{equation}

Our result for $\rho_{\perp}$ is in general agreement with the previous  theory (See e.g. Ref. \cite{10}) For $ gR \gg 1$ our formula   coincides with the  corresponding results of \cite{10,11} for a simple harmonic modulation. Apart from $\nu =1/2$ when the parameter $ gR $ is of the order of or smaller 
than unity (i.e. the radius of the CF cyclotron orbit is of the order of or smaller than the modulation wavelength), we have to modify our results for the magnetoresistivity component $\rho_{\perp}$. The reason is that our approximation for $\rho_{\perp}$  is valid within the relaxation time approximation 
in the Boltzmann equation with relaxation towards the total instead of the local equilibrium distribution function (See Ref. \cite{10}). For $ gR \sim 1 $ the difference between the total and local distribution functions becomes significant. It follows from the results of a detailed and systematic analysis [10,13,14] that for $ gR\sim  1 $ we have to modify the correction term arising due to the modulations in the expression for $\rho_{\perp} $. The function $ J_{1}^2(gR) $ has to be replaced by $ J_{1}^2(gR)/(1-J_{0}^2(gR)) $. This denominator originates from the back scattering term in the collision operator in the Boltzmann equation which describes relaxation towards the local equilibrium. Thus it is the 
consequence of the continuity equation for CFs.

It was shown in Ref. \cite{10} that the expression for the magnetoresistivity  component $ \rho_{\parallel} $ is not changed if we assume the relaxation is towards either local or total equilibrium. Thus we can use our semiquantitative expression (15) for $ \rho_{\parallel} $ within the range of strong effective magnetic fields when $ gR < 1 $. We remark that our result (15) for $ \rho_{\parallel} $  disagrees with the previous theory presented in Refs. \cite{10,13,14,15}. It was concluded in these papers that magnetic modulations themselves cannot influence this component of 
the magnetoresistivity. This discrepancy originates from the difference in the procedure of calculation of the response functions averaged over the period of modulations. Here we first introduce the average conductivity [See Eqs. (6) and (7)] and after that we convert it to the  magnetoresistivity tensor, in contrast to Refs. \cite{10,13,14,15} where the average
is taken last. It enables us to 
keep corrections proportional to $ \displaystyle{\left({\Delta 
\Omega}/{\Omega}\right)^2J_{0}^2(gR)} $.

These corrections are missed when we first calculate the resistivity tensor and then  average it over the period  of modulation as in Ref. \cite{10}. Another source of the discrepancy is the  diffusion correction included to $ \sigma_{xx} $. However this term has a 
clear physical sense and originates from the extra current along the $"x" $ direction due to the guiding center drift.

For $ gR < 1 $ the correction to the magnetoresistivity component $ \rho_{\perp} $ takes on values of the order of unity and increases upon increase of the effective magnetic field. This corresponds to a minimum in the magnetic field dependence of $ \rho_{\perp} $ around $ \nu =1/2 $. 
Such a minimum was observed in experiments \cite{1,12}. The magnetic field dependence of the resistivity component $ \rho_{\parallel} $ for $gR \sim 1$ is more complicated. Within a certain range of $gR$ we may observe a maximum of $ \rho_{\parallel} $ about $ \nu =1/2 $. The maximum can be
developed when the CF mean free path and the modulation period are large enough to obey the inequalities $ l> R $ and $gR < 1 $ for sufficiently small $ B_{eff}. $  Within another
interval of $gR$ this maximum can be converted to a minimum and for some values of $gR$ this component of the magnetoresistivity will exhibit only a very weak dependence on the effective magnetic field. The character of this dependence is determined by the relation between Bessel functions 
$J_{0}^2(gR)$ and $J_{1}^2(gR)$ included in the second term of our expression (15). 

We conjecture that by changing the modulation wavelength $\lambda $ we can observe all three kinds of the dependence of $\rho_{\parallel} $ upon the magnetic field for the same experimental sample. Actually a very weak dependence of the resistivity component $\rho_{\parallel}$ on the effective magnetic field near $\nu =1/2 $ was observed in the experiments
of Ref. \cite{12} and a maximum was observed in the experiments of Ref. \cite{1}. In both cases the resistivity component $\rho_{\perp} $  exhibits a large maximum about $ \nu =1/2 $.

\begin{figure}[t]
\begin{center}
\includegraphics[width=5cm,height=11cm]{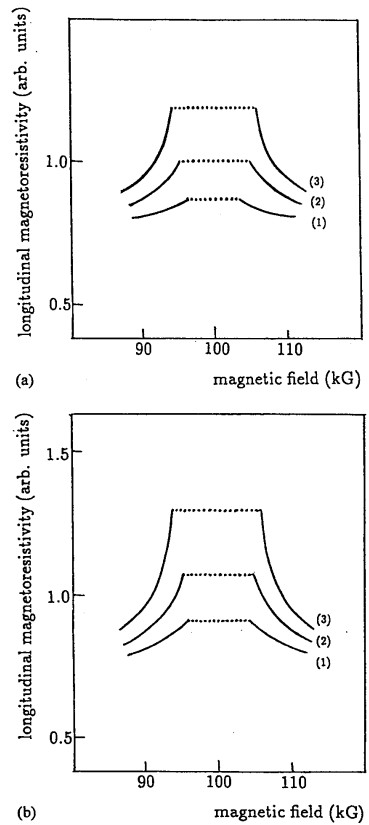}
\caption{
The magnetic field dependence of the longitudinal dc resistivity near $ \nu = 1/2$ for a circular (a) and elliptic (b) CF-FS for different magnitudes of density modulations. Curves are plotted for  $ l \sim 10^{-4}$ cm, $\lambda = 1.3 \mu$m, $ \Delta n /n = 0.03 (1);\ 0.05 (2); \ 0.07 (3)$.  The dotted part of the theoretical curve corresponds to the region of the values of $ B_{eff} $ where Eq. (15)
cannot be applied.
}   
\label{rateI}
\end{center}
\end{figure}

\begin{figure}[t]
\begin{center}
\includegraphics[width=4.5cm,height=5cm]{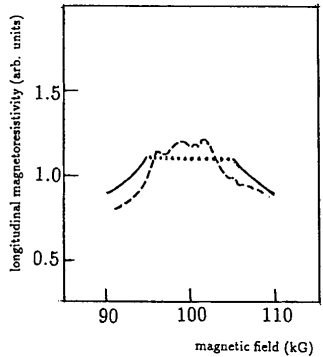}
\caption{
dc magnetoresistivity versus the effective magnetic
field. Dashed line -- experiment of Ref. \cite{1}, solid line -- theory for the parameters $  n = 1.1 \times 10^{11}$ cm$^{-1}$, $\Delta n /n = 0.03$ , $ \lambda = 1.5 \mu $m , $ l = 10^{-4 }$cm.  The dotted part of the theoretical curve corresponds to the region of the values of $ B_{eff} $ where Eq. (15)
cannot be applied.
}   
\label{rateI}
\end{center}
\end{figure}

We tried to describe the maximum in the magnetic field dependence of the longitudinal magnetoresistivity about $\nu =1/2 $ using our result (15) for the parameters close to the parameters of the corresponding experiments of \cite{1}: $ n \sim 1.1\times 10^{11}$ cm$^{-1},\ p_{F}\sim
1.2\times10^{-21} $g cm/s$^2,\ \lambda\sim 1.5\mu $m,$\ l\sim 10^{-4}$ cm and we obtained a maximum in the longitudinal magnetoresistivity near $\nu =1/2$.  The magnitude of the this maximum strongly  depends on the magnitude of the density modulations [see Fig. 1(a)]. However for a circular CF-FS the maximum about $ \nu = 1/2 $ is noticeably smaller in magnitude than the experimental results of \cite{1} for reasonable values of $ \Delta n/n. $  We can reach much better agreement with this experiment for a noncircular CF-FS. In principle, a noncircular shape of the CF-FS may originate from the effect of the crystalline field of adjacent layers of GaAS/AlGaAs. For clean samples like those used in the experiments of \cite{1}this effect can be noticeable. Suppose that the CF-FS
is not a circle but an ellipse. In this case the parameter 
$ gR = \displaystyle{{gp_{1}c}/{|e|B_{eff}}} \ $  
includes the maximum value of the CF quasimomentum component $ p_{x} $ which does not coincide with $p_{F}$. This can give us a noticeable enhancement in the magnitude of the maximum as is shown in Fig. 1(b). Plotting the curves in this figure we assumed that $ p_1 $ is $ \sim 1.0 \times 10^{-21} $ g cm/s$^2. $ Using this value of $ p_1 $ we get reasonable agreement with the experimental data of Willett et al as is shown on Fig. 2.

Our simplified formula (20) cannot be applied to the region immediately adjacent to $ \nu = 1/2 $ where the condition $ \Delta B / B_{eff} \ll 1 $ is not satisfied. So we cannot analyze the effect of channeled orbits of CFs which occur immediately adjacent to $ \nu = 1/2 $ where the effective magnetic field is of the same order of magnitude as
the oscillating  correction $ \Delta B. $ To analyze the dc response of the modulated 2DEG for $ \Delta B \simeq B_{eff} $ we have to solve the CF transport problem consistently for example following the way treated in Ref. \cite{10}. 
Nevertheless our semiquantitative approach captures the essential physics of the considered effect and gives simple analytical results applicable for the comparison with experimental data.

In summary, we show that within the single relaxation time approximation we obtain good semiquantitative agreement with experimental observation of a maximum in parallel resistivity $ \rho_{\parallel} $, observed in the experiments of Ref. \cite{1}. We predict that the magnetic field dependence of
this component of the magnetoresistivity can be of different character for different $gR  $ i.e by changing the wavelength of modulation the maximum in $\rho_{\parallel} $ can be converted to a minimum or to a negligeably weak dependence of this resistivity component upon the effective magnetic
field. This maximum as well as the minimum in the magnetic field
dependence of the resistivity $\rho_{\perp} $ about $\nu =1/2 $ arises due to the inhomogeneity of the effective magnetic field in a modulated 2DEG near one half filling.

We thank W.R. Willett for kindly giving us a preprint of reference \cite{1} and G.M. Zimbovsky for help with the manuscript.  We also thank S. Das Sarma for discussions of possible asymmetry effects on the CF-FS due to the crystal field present on the ultra clear surfaces of AlGaAs, Used in the magnetotransport experiments. Support from a PSC-CUNY FRAP "In -- Service" Award is acknowledged.


\begin{thebibliography}{99}
\bibitem{23} An alternative theoretical description of quantum Hall system at $ \nu = 1/2 $ was recently proposed (see Refs. \cite{18,19,20}). In the framework of this approach the relevant quasiparticles of the system are neutral dipolar fermionic objects. However a magneto-transport theory
based on this new physical picture \cite{21} gives similar results for observables.
\end{thebibliography}
\end{document}